\documentstyle[12pt]{article}

\def\tspace{\baselineskip = .35in}
\begin{document}
\begin{titlepage}
\begin{flushright}
        BA-97-53\\
        FERMILAB--PUB--97/382--T\\
        December 1997\\[0.5in]
\end{flushright}
\begin{center}
{\Large Fermion masses in $SO(10)$ with a single adjoint Higgs field\\[1in]}

{\bf Carl H. ALBRIGHT}\\[0.1in]
Department of Physics, Northern Illinois University, DeKalb, Illinois
60115\footnote{Permanent address}\\
       and\\
Fermi National Accelerator Laboratory, P.O. Box 500, Batavia, Illinois
60510\footnote{Electronic address: albright@fnal.gov}\\[0.4in]

{\bf S.M. BARR}\\[0.1in]
Bartol Research Institute,
University of Delaware, Newark, DE 19716\footnote{Electronic address:
smbarr@bartol.udel.edu}\\[0.6in]

\begin{abstract}

It has recently been shown how to break $SO(10)$ down to 
the Standard Model in a realistic way with only one adjoint Higgs. 
The expectation value of this adjoint must point in the $B-L$ direction.
This has consequences for the possible form of the quark and lepton mass
matrices. These consequences are explored in this paper, and it is found
that one is naturally led to consider a particular form for the masses
of the heavier generations. This form implies typically that there should 
be large (nearly maximal) mixing of the $\mu$ and $\tau$ neutrinos.
An explanation that does not involve large $\tan \beta$ also emerges for 
the fact that $b$ and $\tau$ are light compared to the top quark.

\end{abstract}
\begin{flushleft}
PACS numbers: 12.15.Ff, 12.10.Dm, 12.60.Jv, 14.60.Pq \hfill 
	e-Print Achive: hep-ph/9712488
\end{flushleft}
\end{center}
\end{titlepage}
\tspace
\setcounter{page}{2}
\section{Introduction}

\qquad For a number of reasons, $SO(10)$ is widely considered to be the 
most 
attractive grand unified group. It achieves complete quark-lepton 
unification 
for each family; explains the existence of right-handed neutrinos and of 
``seesaw" neutrino masses; has certain advantages for baryogenesis, in 
particular, since $B-L$ is broken \cite{broken}; and has the greatest 
promise for explaining the pattern of quark and lepton masses \cite{ramond}
- \cite{a-n}. Some progress has 
been made in constructing $SO(10)$ models in superstring theory, it now being
known that there are perturbative ground states of the heterotic string 
with three generations of quarks and leptons \cite{dienes review}.

It has been shown that there are limitations in the context of
perturbative superstring theory on supersymmetric grand unified
models which have more than a single adjoint Higgs field. 
In particular, it had been argued that if there are multiple
adjoints in realistic models they must have the same charges
under local symmetries. (They may have different discrete
gauge charges, however.) This makes it significantly harder
to construct realistic models in which there are several 
adjoints which couple in different ways \cite{ss adj}. On the
other hand, until recently, it was not known how to break 
$SO(10)$ without either using three adjoint Higgs fields 
\cite{3 adj} or having colored pseudo-goldstone fields 
that largely vitiated the unification of gauge couplings 
\cite{his, b-b color}. However, in a recent paper \cite{b-r},
a satisfactory mechanism was proposed for achieving natural breaking
of $SO(10)$ without more than one adjoint Higgs field. But in that
paper only the Higgs sector was considered. This raises the question of
whether quarks and leptons can be incorporated in a satisfactory way into
models which employ that mechanism of symmetry breaking. 

There are two aspects to this question. First, it is not obvious whether
a single adjoint Higgs is sufficient to give a realistic pattern of quark
and lepton masses. If there is only one adjoint Higgs field in $SO(10)$, 
its vacuum expectation value must point in the $B-L$ direction in order 
to 
produce the doublet-triplet splitting \cite{dim-wil}. This greatly 
constrains the
possibilities for the quark and lepton masses, as this adjoint VEV is the 
only one that breaks the $SU(5)$ subgroup of $SO(10)$ at the unification 
scale, and therefore the only one that can break the ``bad" $SU(5)$ 
relations 
such as $m^0_{\mu} = m^0_s$. (The superscript `$0$' refers throughout to 
parameters at the unification scale.) All models in the literature which
attempt to explain the pattern of fermion masses in the context of $SO(10)$
make use of adjoint VEVs that point in directions other than $B-L$ \cite{
ramond} - \cite{a-n}. 

The second issue has to do with the stability of the gauge hierarchy.
In $SO(10)$, as in any unified model, there are higher-dimension
operators that would destabilize the hierarchy, and which must therefore
be forbidden by some local symmetry or other principle. These local
symmetries constrain the possible couplings of the Higgs fields and therefore
the possible Yukawa couplings of the quarks and leptons. Conversely,
the existence of realistic quark and lepton Yukawa interactions
may be incompatible with any symmetry that could stabilize the hierarchy,
and may therefore imply the presence (because of Planck-scale effects) of
operators that destroy the hierarchy.

In this paper we show that a realistic pattern of quark and lepton masses
can be achieved in a natural way using only one adjoint Higgs and the 
mechanism
for symmetry-breaking proposed in \cite{b-r}. We find, indeed, that the
possibilities are tightly constrained, and under certain reasonable
requirements the basic structure that we find may be unique. This structure
is fairly simple: it does not require that there be any Higgs fields or
any symmetries beyond those introduced in \cite{b-r} to achieve $SO(10)$ 
breaking to $SU(3) \times SU(2) \times U(1)$. It also provides an explanation
of many of the qualitative and quantitative features of the quark and
lepton masses and mixings. 

There are two interesting features of the structure
to which we are led. First, it typically gives large, and indeed nearly
maximal, mixing of $\nu_{\mu}$ with $\nu_{\tau}$. This is possibly of great
significance in light of the evidence of such mixing coming from
atmospheric neutrino observations. Second, an interesting explanation
emerges of the smallness of $m_b$ and $m_{\tau}$ 
compared to $m_t$ that does not involve large $\tan \beta$.

\section{Review of the Breaking of SO(10)}

\qquad Before turning to the problem of quark and lepton masses, let us 
briefly
review the mechanism proposed in \cite{b-r} for breaking $SO(10)$ with only
a single adjoint. The Higgs superpotential has the form
\begin{equation}
W = T_1 A T_2 + M_T T_2^2 + W_A + W_C + W_{CA} + W_{TC},
\end{equation}

\noindent
where $T_1$ and $T_2$ are ${\bf 10}$'s and $A$ is a ${\bf 45}$. $W_A$ is
a set of terms that produces the ``Dimopoulos-Wilczek" form for the
expectation value of $A$: $\langle A \rangle = {\rm diag}(0,0,a,a,a)
\times i \tau_2$, where $a \sim M_G$. This is equivalent to saying that
the VEV of $A$ is proportional to the generator $B-L$. This form for 
$\langle A \rangle$ couples the 
color-triplets in $T_1$ amd $T_2$, but not the weak-doublets. The effect
of the first two terms in Eq. 1 is to give superheavy masses to all the
color triplets in $T_i$ but leave the pair of weak-doublets in $T_1$ light.
The simplest form for $W_A$ that works is
\begin{equation}
W_A = {\rm tr} A^4/M + M_A {\rm tr} A^2.
\end{equation}

\noindent Here and in the following, all explicit denominator masses are 
regarded as Plank scale masses, i.e., $M_P$.

To break $SO(10)$ completely to the Standard Model requires also
Higgs in the spinor representation which must get vacuum expectation
values in the $SU(5)$-singlet direction. If $C$ and $\overline{C}$
are respectively a ${\bf 16}$ and ${\bf \overline{16}}$, then a simple
form for $W_C$ is 
\begin{equation}
W_C = X(\overline{C} C)^2/M_C^2 + f(X),
\end{equation}

\noindent
where $X$ is a singlet field, and $f(X)$ is a polynomial in $X$ that
has at least a linear term. Then the f-flat condition $F_X = 0$ forces 
$C$ and 
$\overline{C}$ to get VEVs.  

The terms $W_{CA}$ couple the spinor sector
($C$, $\overline{C}$) to the adjoint sector ($A$). This is necessary 
\cite{b-r}
to prevent light, color-singlet pseudo-goldstone fields from being
produced by breaking of the unified symmetry. The only mechanism 
known to do this without involving several adjoint fields was
proposed in \cite{b-r}. The form of $W_{CA}$ given there is
\begin{equation}
W_{CA} = \overline{C}'(P A / M_1 + Z_1) C +
\overline{C} (P A/M_2 + Z_2) C'.
\end{equation}

\noindent
Here $C'$ and $\overline{C}'$ are an additional ${\bf 16} + {\bf 
\overline{16}}$ pair, and $P$, $Z_1$ and $Z_2$ are singlets. $C'$ and
$\overline{C}'$ have vanishing VEVs, which ensures that $W_{CA}$
does not destabilize the hierarchy ({\it i.e.} the Dimopoulos-Wilczek
form of $\langle A \rangle$) by contributing to $F_A$. The $F_{C'} =0$
and $F_{\overline{C}'} = 0$ equations lead to the conditions
$(PA/M_1 + Z_1) C = \overline{C}(PA/M_2 + Z_2) = 0$ having a 
discrete number of solutions, for one of which $\langle 
C \rangle$ and $\langle \overline{C} \rangle$ point in the $SU(5)$-singlet
direction. These two equations then fix the relative magnitudes of the
VEVs of the singlets $P$ and $Z_i$. There is one linear combination
of these singlets that is not fixed by the terms in Eq. 1, but this can
be fixed by radiative effects after supersymmetry breaks \cite{b-r}.

Finally, the $W_{TC}$ term which was not included in \cite{b-r} is added 
here 
in order to induce an electroweak-breaking VEV in the spinor $C'$.  This 
VEV 
will help to generate the desired texture in the fermion mass matrices.
For this purpose we set 
\begin{equation}
        W_{TC} = \lambda T_1\overline{CC} 
\end{equation}

\noindent where $\lambda$ is a dimensionless coefficient which, as we 
shall 
see later, must be somewhat smaller than one --- about $1/20$.  From the
$F_{\overline{C}}^* = 0$ equation, which gives 
\begin{equation}
0 = 2 \lambda T_1 \overline{C} + (P A/M_2 + Z_2)C'.
\end{equation}

\noindent it then follows that since $\overline{C}$, $P$, $A$, and $Z_i$ 
all 
have superlarge VEVs in the $SU(5)$ {\bf 1} direction, while the Higgs 
doublets of $T_1$ are assumed to develop weak-scale VEV's in the $SU(5)$
{\bf 5} and ${\bf \overline{5}}$ directions, the $SU(2)_L$-doublet in 
$C'$ must 
also develop a weak-scale VEV in the $SU(5)\ {\bf \overline{5}}$ 
direction. 

This set of terms gives a complete breaking of $SO(10)$ down to the
Standard Model group without fine-tuning of parameters and without
pseudo-goldstone fields. The mass $M_T$ appearing in Eq. (1) must
arise from the expectation value of some field or product of fields.
Two viable possibilities are $P^2$ and $Z_i$. 

The stability of the hierarchy requires that certain types of
higher-dimension terms not arise, in particular terms that give
effectively $T_1^2$, $\overline{C} A C$, $\overline{C}C{A^2}/M$ , or 
$Z_i^n$. 
The first of these, $T_1^2$, would directly give superheavy mass to the 
doublet 
Higgs fields.  Both $\overline{C} A C$ and $\overline{C}C{A^2}/M$ would 
destabilize the Dimopoulos-Wilczek form of $\langle A \rangle$; hence the 
choice of a higher order term in the $W_{C}$ superpotential of (3). The 
appearance of  $Z_i^n$ would cause a conflict between the
$F_{Z_i} = 0$ equations and the $F_{C'} = 0$ and $F_{\overline{C}'}
= 0$ equations. In \cite{b-r} it was shown that a simple $U(1) \times 
Z_2 \times Z_2$ symmetry is sufficient to rule out all dangerous 
operators. 
In order to obtain the desired appearance of the $\lambda T_1 
\overline{CC}$ 
term in $W_{TC}$ along with the rest of the Higgs superpotential, the 
$U(1) 
\times Z_2 \times Z_2$ charges are reassigned as follows:
\begin{equation}
\begin{array}{rllll}
        & A(0^{+-}),\quad & T_1(1^{++}),\quad & T_2(-1^{+-}) \\[0.1in]
        & C({1\over{2}}^{-+}),\quad & \overline{C}(-{1\over{2}}^{++}),\quad
                & C'(\left[{1\over{2}}-p\right]^{++}),\quad & \overline{C}'
                (\left[-{1\over{2}}-p\right]^{-+}) \\[0.1in]
        & X(0^{++}),\quad & P(p^{+-}),\quad & Z_1(p^{++}),\quad & 
Z_2(p^{++})\\
        \end{array}
\end{equation}

\section{B - L Generator and Fermion Mass Matrix Textures}

\qquad We have succeeded in constructing a simple superpotential for the
quark and lepton fields that gives the fermions realistic masses
and makes use of no Higgs superfields beyond the set found necessary
to achieve a satisfactory breaking of $SO(10)$ in \cite{b-r}, namely
$T_i$, $A$, $C$, $\overline{C}$, $C'$, $\overline{C}'$, and the
singlets $X,\ P,\ Z_1$ and $Z_2$. To help understand this superpotential 
before writing it down, we explain the kind of textures that
are needed if only one adjoint is available with its VEV in the $B-L$ 
direction. The desired textures for the mass matrices $U$, $D$, and
$L$ are of the form
\begin{equation}
U \cong \left[ \begin{array}{ccc} 0 & 0 & 0 \\
0 & 0 & F/3 \\ 0 & -F/3 & E \end{array} \right] v_u,
\end{equation}

\begin{equation}
D \cong \left[ \begin{array}{ccc} 0 & 0 & G' \\
0 & 0 & F/3 + G \\ 0 & -F/3 & E \end{array} \right] v_d,
\end{equation}

\noindent and
\begin{equation}
L \cong \left[ \begin{array}{ccc} 0 & 0 & 0 \\
0 & 0 & -F \\ G' & F + G & E \end{array} \right] v_d.
\end{equation}
   
\noindent These matrices are written so that the left-handed antifermions
multiply them from the left and the left-handed fermions from the right.
We imagine that some of the zero entries in the first row and column
actually get small contributions from higher order terms so that the
first generation will not remain exactly massless. This will
be discussed later.
Note that the parameter $F$ is multiplied by a factor of $B-L$ 
everywhere. 
Suppose that we assume that
$G \sim E \gg F$. Denote the small parameter $F/E$ by the symbol
$\epsilon$, and the $O(1)$ parameter $\sqrt{G^2 + G'^2}/E$ by $\rho$.
Then it is easy to see that the following relations hold: 
\begin{equation}
\begin{array}{l}
m^0_c/m^0_t \cong \epsilon^2/9, \\[0.1in]
m^0_s/m^0_b \cong \epsilon \rho/3(1 + \rho^2) \sim \epsilon/3, \\[0.1in]
m^0_{\mu}/m^0_{\tau} \cong \epsilon \rho/(1 + \rho^2) \sim \epsilon,\\[0.1in]
m^0_{\tau} \cong m^0_b, \qquad m^0_{\mu}/m^0_s \cong 3,\\[0.1in]
V_{cb} \cong \epsilon \rho^2/3(1 + \rho^2) \sim \epsilon/3.\\
\end{array}
\end{equation}

Thus the following facts would be explained: the equality at the GUT scale
of the $b$ and $\tau$ masses, the Georgi-Jarlskog factor of 3 between
the $\mu$ and $s$ masses at the GUT scale \cite{g-j}, why $V_{cb}$ is of
order $m_s/m_b$, why $m_c/m_t$ is much smaller than both $m_s/m_b$
and $m_{\mu}/m_{\tau}$, and why the second generation masses are
small compared to the third, and the first generation masses are very small
compared to the second. This list contains most of the salient features 
of 
the quark and lepton spectrum. It is important to note how some of these
relations are achieved, and therefore the rationale for the form of the
textures. 

In our model the only generator of $SO(10)$ available for constructing
the textures is $B-L$. As we shall see, it is a simpler matter for this 
generator to appear in the off-diagonal entries than in the diagonal ones.
However, if the 23 and 32 entries are just proportional to $B-L$, while the
33 entries are proportional to the identity, then the ratio
$(m_{\mu}/m_{\tau})/(m_s/m_b)$ is 9 instead of the Georgi-Jarlskog
value of 3. It is therefore essential to have asymmetrical entries 
like those denoted by $G$ and $G'$. With $G$ or $G'$ being much larger
than $F$ and {\it not} depending on $B-L$, the 
desired ratio of 3 for $m^0_{\mu}/m^0_s$ is obtained. As we will
see, such asymmetrical entries can be achieved simply by integrating out
$SO(10)$ ${\bf 10}$'s of fermions, since these contain $SU(5)$ 
${\bf \overline{5}} + {\bf 5}$ (which contain $d^c_L$ and $l_L$) but 
not $SU(5)$ ${\bf 10}$ (which contain $d_L$ and $l^c_L$). 
Moreover, entries produced in this way will appear only in the down quark 
and charged lepton mass matrices, $D$ and
$L$; but not in the up quark and Dirac neutrino mass matrices, $U$ or $N$.
(This follows from the fact that they come from effective operators 
of the form ${\bf 16} {\bf 16} {\bf 16}_H {\bf 16}_H$, where ${\bf 16}_H$
contains the ${\bf \overline{5}}$ but not the ${\bf 5}$ of $SU(5)$.)
This then automatically explains why the ratio $m_c/m_t$ is much smaller 
than the $m_s/m_b$ and 
$m_{\mu}/m_{\tau}$ ratios. The fact that the entries $G$ and $G'$
appear in $D$ but not in $U$ also explains why $V_{cb}$ does not
vanish. (Of course, $V_{cb} = 0$ is a minimal $SO(10)$ relation.)

\section{Important Conclusion about Neutrino Mixing}

\qquad Careful consideration of those possibilities available that use only
the generator $B-L$ leads to the conclusion that the textures given above
are likely to be the only ones that satisfy the requirements of simplicity
and realism. Other structures tend to be more complicated, or require
artificial numerical relationships among parameters to reproduce the
qualitative and quantitative features of the spectrum of quarks and
leptons. 

These textures already have an interesting phenomenological
consequence, namely, that they predict large mixing of $\nu_{\mu}$
and $\nu_{\tau}$. The neutrino mixing angles arise from the mismatch 
between the unitary transformations required to diagonalize the charged
leptom mass matrix, $L$, and the neutrino mass matrix, $M_{\nu}$.
The neutrino mass matrix can be written in the familiar seesaw
form: $M_{\nu} = - N^T M_R^{-1} N$, where $M_R$ is the superheavy
Majorana mass matrix of the right-handed neutrinos, and $N$ is the
Dirac mass matrix for the neutrinos. Little can be said at present about 
the form of $M_R$ as there are many possible ways that the right-handed
neutrinos can get mass. However, the form of $N$ is closely connected
to the forms of $U$, $D$, and $L$. In fact, given the forms shown
in Eqs. (8) - (10), one expects $N$ to have the form
\begin{equation}
N = \left( \begin{array}{ccc} 0 & 0 & 0 \\ 0 & 0 & -F \\
0 & F & E \end{array} \right).
\end{equation}

\noindent
Precisely this form will indeed arise from the superpotential
that we shall discuss in the next section.
The similarity of structure of $N$ and $U$ is a typical feature of
$SU(5)$ and $SO(10)$ models. The difference in the coefficient of the
$F$ term is, of course, just due to the generator $B-L$. The 
$G$ and $G'$ terms are absent from $N$ just as they are from $U$ for the
reasons explained above. 

One sees immediately that the $13$ and
$23$ angles required to diagonalize $M_{\nu}$ vanish in the limit
that the second generation masses go to zero ({\it i.e.} $F/E 
\equiv \epsilon \longrightarrow 0$) and the first generation masses
go to zero, no matter what the form of $M_R$. Nevertheless, it is 
possible that 
the texture of $M_R$ is such that
these angles are numerically large in spite of being formally of 
order $\epsilon$. However, we will assume that $M_R$ does not have such
a special form, and therefore that one can neglect these angles.
With this plausible assumption, the mixing angle between $\nu_{\mu}$
and $\nu_{\tau}$ can be read off directly from the matrix $L$.
It is given by $\tan \theta_{\mu \tau} \cong \sqrt{G^2 + G^{'2}}/E
= \rho$. One then finds that
\begin{equation}
\tan \theta_{\mu \tau} \cong 
\rho \equiv 3 V_{cb}^0/(m^0_{\mu}/m^0_{\tau}) \cong 1.8.
\end{equation}

\noindent It is quite striking that the constraint of having $SU(5)$ 
broken 
only by an adjoint pointing in the $B-L$ direction, which is in essence
a minimality condition on the Higgs sector, leads in a natural way to
textures for the quark and lepton mass matrices that predict large mixing
of the $\mu$ and $\tau$ neutrinos.  The consequences of this implication for
neutrino mixing will be explored more fully elsewhere.\cite{ab2}

\section{Yukawa Superpotential Yielding the Desired Textures}

\qquad We will now show how these textures arise in a straightforward
way from a few terms in the superpotential. We distinguish the third
generation quarks and leptons, which we denote ${\bf 16}_3$, from
the other two generations, which we denote ${\bf 16}_i$, $i = 1,2$.
In addition, we posit the existence of some ``vectorlike" sets of
quarks and leptons to be ``integrated out", namely ${\bf 16} +
{\bf \overline{16}}$, ${\bf 10}$ and ${\bf 10}'$. The proposed Yukawa 
superpotential has the following form:
\begin{equation}
\begin{array}{ccl}
W_{Yukawa} & = & {\bf 16}_3 {\bf 16}_3 T_1 \\
& & \\
& + & {\bf 16} {\bf \overline{16}} P + {\bf 16}_3 {\bf \overline{16}} A
+ a_i {\bf 16}_i {\bf 16} T_1 \\
& & \\
& + & {\bf 10} {\bf 10}' \overline{C} C/M_P + c_i {\bf 16}_i {\bf 10} C
+ {\bf 16}_3 {\bf 10}' C'.
\end{array}
\end{equation}

\noindent
As in the Higgs superpotential, we have suppressed most of the 
dimensionless coefficients, which are assumed to be of order unity.
However, we have explicitly written the two Yukawa coefficients that
carry the family index $i$, which, of course, is summed over. 
Recall that the Higgs fields $T_1$ and $C'$ each develop weak-scale VEV's,
while $A,\ C,\ \overline{C},\ P,\ Z_1$ and $Z_2$ all acquire superlarge 
VEVs. 
No VEV's appear for $\overline{C'}$ or $X$.

The 33 elements denoted by $E$ in the $U,\ D$ and $L$ matrices of (8) - 
(10) 
obviously arise directly from the first term in Eq. (14) as 
illustrated in Fig. 1(a).
The $F$ contributions to the matrix elements arise from the next three
terms in Eq. (14), which contain the spinors ${\bf 16}$ and ${\bf 
\overline{16}}$. This is easiest to see diagrammatically by
considering Fig. 1(b). By integrating out those spinors one effectively
obtains a term of the form $a_i {\bf 16}_i {\bf 16}_3 \langle A \rangle  
\langle T_1 \rangle/M_G$. Because the vacuum expectation value of $A$
is proportional to $B-L$, this term will have a factor of $B-L$
of the field contained in ${\bf 16}_3$ (or, equivalently, $-(B-L)$
of the field contained in ${\bf 16}_i$). Without loss of generality,
one can take the Yukawa coefficient $a_i$ to point in the 2 direction.
Thus one has $F [(B-L)_{f} f^c_2 f_3 + (B-L)_{f^c} f^c_3 f_2] \langle T_1
\rangle$, where $F$ is a dimensionless combination of VEVs and
Yukawa couplings. This form also explains why it is hard for the generator
$B-L$ to appear in a diagonal element of the mass matrices, for the 
combination
$[(B-L)_f + (B-L)_{f^c}] f^c_i f_i$ vanishes for the diagonal ii matrix
element.

The $G$ and $G'$ contributions to the mass matrices in (8) - (10) arise from
the last three terms in Eq. (14), which contain the vector fields ${\bf 10}$
and ${\bf 10'}$ as can be seen diagrammatically from Fig. 1(c). 
Having defined the 2 direction to be that of $a_i$, there is
no freedom left, and $c_i$ will have components in both the 1 and 2
directions. Since as noted earlier, the VEV's of $C$ and $C'$ point 
respectively in the ${\bf 1}$ and ${\bf \overline{5}}$ $SU(5)$ directions,
it is clear that only the ${\bf \overline{5}}
({\bf 16}_i){\bf 5}({\bf 10}) \langle {\bf 1}(C) \rangle$ and ${\bf 
\overline{10}}({\bf 16}_3) {\bf \overline{5}}({\bf 10'}) \langle {\bf 
\overline{5}}(C') \rangle$ components of the last two terms in the 
superpotential of (14) can contribute to the mass diagram in Fig. 1(c).  
(Here 
and throughout ${\bf p}({\bf q})$ denotes an $SU(5)$ ${\bf p}$ contained 
in 
an $SO(10)$ ${\bf q}$.)
Hence with the convention that the mass matrices are to be multiplied from
the left by left-handed antifermions and from the right by left-handed 
fermions, the diagram depicted in Fig. 1(c) can only contribute to the 
13 and 23 elements of the down quark mass matrix $D$ and the 31 and 32 
elements of the charged lepton mass matrix $L$.  The up quark mass matrix
$U$ and the Dirac neutrino mass matrix $N$ receive no such contributions.

One can also easily see the origin of the $G$ and $G'$ terms 
directly from the superpotential terms in Eq. (14). 
The ${\bf 5}({\bf 10})$ has a mass term with the linear combination
of superfields $\langle \overline{C} C/M_P \rangle {\bf \overline{5}}
({\bf 10}') +
c_i \langle C \rangle {\bf \overline{5}}({\bf 16}_i)$. But this linear
combination lies nearly exactly in the $c_i {\bf 16}_i$ direction,
because of the $M_P^{-1}$ Planck scale suppression factor. Thus ${\bf 
\overline{5}}({\bf 10}')$ is almost purely one of the light ({\it i.e.} 
weak-scale) multiplets, and in generation space points partly in the 1 
and 
partly in the 2 directions. It then follows directly that the term
${\bf 16}_3 {\bf 10}' C'$ gives the $G$ and $G'$ entries. Note 
that direct calculation of the mass matrix elements shows these entries 
are 
not suppressed by powers of $M_P$ as one might naively think from Fig. 1(c).

Before turning to the question of how the small first generation masses
arise, we note that the terms in the Yukawa superpotential of $13$ do not 
destabilize the gauge hierarchy. With the assignments given in (7) for 
the 
Higgs multiplets, the charges of the chiral multiplets are completely 
determined by the terms appearing in (14):
\begin{equation}
\begin{array}{ll}
        {\bf 16}_3(-{1\over{2}}^{++}), \qquad & {\bf 16}_i(\left[-{1\over{2}}
                + p\right]^{++}),\qquad i = 1,2\\[0.1in]
        {\bf 16}(-{1\over{2}}^{++}), \qquad & {\bf \overline{16}}
                ({1\over{2}}^{++})\\[0.1in]
        {\bf 10}(-p^{-+}), \qquad & {\bf 10'}(p^{++})\\
\end{array}
\end{equation}

\noindent
The value of the charge $p$ depends on which field or fields couple 
to $T_2^2$. Two viable choices are $p=1$ or $p = 2$, giving respectively
that the mass term for $T_2$ is of the form $T_2^2 P^2/M_P$ or
$T_2^2 Z_i$. It is easily checked that the $U(1) \times Z_2 \times
Z_2$ forbids any destabilizing terms, such as those containing
factors of $T_1^2$, $\overline{C} A C$, and $Z_i^n$ as discussed in the 
pure Higgs field case.  There are some
higher-dimension terms not included in Eq. (14) that are allowed by the 
symmetry, such as ${\bf 10}^2 P^2/M_P$, but these prove to be harmless.

The requirement of stability of the gauge hierarchy does dictate
an important feature of the structure of the Yukawa superpotential in 
(14), 
namely that $C'$ acquires a weak-scale ${\bf \overline{5}}({\bf 16})$ 
$SU(2)_L 
\times U(1)_Y$-breaking VEV, and that $C'$ and $T_1$ therefore mix. One 
might 
imagine that the $G$ and $G'$ terms in the matrices of (8) - (10) could 
be 
generated without a spinor Higgs field acquiring an $SU(2)_L
\times U(1)_Y$-breaking VEV. This could happen via the diagram in Fig. 2,
if instead of the terms in Eq. (14) there were the following terms:
${\bf 16}_3 {\bf 16}_3 T_1 + {\bf 16} {\bf \overline{16}} P +
a_i {\bf 16}_i {\bf \overline{16}} A + {\bf 16}_3 {\bf 16} T_1
+ {\bf 10} {\bf 10} S + c_i {\bf 16}_i {\bf 10} C + {\bf \overline{16}}
{\bf 10} \overline{C}$. However, it is easy to see that the
existence of the terms ${\bf 10} {\bf 10} S$, ${\bf 16}_i {\bf 10} C$,
${\bf \overline{16}} {\bf 10} \overline{C}$, ${\bf 16}_i {\bf 
\overline{16}} A$, and $X(\overline{C} C)^2/M^2_P$ would imply that the
term $\overline{C} A C S/M_P$ is allowed by the symmetry;
this term would destroy the gauge hierarchy and such a form for the 
Yukawa superpotential is unacceptable for the doublet-triplet splitting
solution.

Thus it seems that generating simple and realistic textures for the
quark and lepton mass matrices requires that $C'$ break the electroweak
symmetry and mix with $T_1$. This is an important fact, for 
it may also hold the key to explaining why $t$ is much heavier
than $b$ and $\tau$, which is otherwise somewhat mysterious in the
context of $SO(10)$. This point can be seen from Eq. (6), which says that
the linear combination of ${\bf \overline{5}}(T_1)\cos \theta  
+ {\bf \overline{5}}(C')\sin \theta$, where $\tan \theta =
\langle (P A/M_2 + Z_2) \rangle / (2 \lambda) \langle \overline{C}
\rangle$, has a vanishing VEV. In fact, from
the term $\left| F_{\overline{C}} \right|^2$ in the scalar potential,
it is clear that this linear combination is superheavy. The orthogonal
linear combination is the field $H'$ of the MSSM, while $H$ has the usual
definition:
\begin{equation}
\begin{array}{l}
H' = {\bf \overline{5}}(C')\cos \theta  - {\bf \overline{5}}(T_1) \sin 
\theta\\
H = {\bf 5}(T_1).
\end{array}
\end{equation}

\noindent
Therefore the ratio of the $b$ to $t$ masses is determined by the angle 
$\theta$, in particular:
\begin{equation}
m^0_b/m^0_t = m^0_{\tau}/m^0_t = \sin \theta (\langle H' \rangle/
\langle H \rangle) = \sin \theta/ \tan \beta.
\end{equation}

\noindent
But from the fact that $\langle P \rangle \sim \langle A \rangle
\sim \langle \overline{C} \rangle \sim M_G$, while $\langle
Z_i \rangle \sim M_G^2/M_P$, one finds $\tan \theta \sim
\lambda^{-1} M_G/M_P$. Therefore, the smallness of the mass ratios
in Eq. (17) may be due to small $\sin \theta$ rather than
large $\tan \beta$. The authors of \cite{cdrw} pursued a similar attempt
to lower $\tan \beta$ by reducing the ratio of the bottom to top Yukawa 
couplings in $SO(10)$ models.  Here with $\lambda \sim 1/20$ the correct
mass ratios are obtained with $\tan \beta \sim 1$. This would alleviate
the problem of Higgsino-mediated proton-decay, the amplitude
for which is proportional to $\tan \beta$ for the large $\tan \beta$ case.
To suppress Higgsino-mediated proton decay then requires that $M_T$
(see Eq. 1) be made small compared to $M_G$. This, however, tends
to increase $\alpha_s$. Thus, the problems of $SO(10)$ are alleviated
if $\tan \beta$ is small.

So far we have not specified how the quarks and leptons of the first
generation get masses. There are a number of possibilities, all of
which require integrating out additional vectorlike quark/lepton
representations to get effective higher-dimensional Yukawa operators.
One such effective operator is 
\begin{equation}
W' = {\bf 16}_i {\bf 16}_j \overline{C}^{\dag} C' Z_k^{\dag}.
\end{equation}

\noindent
This operator can be obtained by integrating out the vectorlike
representations ${\bf 16}',\ {\bf \overline{16}'},\ {\bf 10}''$ and 
${\bf 10}'''$, as shown in Fig. 3. This operator contributes only
to $D$ and $L$, and thus explains why $m_u/m_t \ll m_d/m_b, m_e/m_{\tau}$.
The $U(1) \times Z_2 \times Z_2$ charges of these additional vectorlike
representations can be read off from Fig. 3, using the charges that
have already been given. It is straightforward to show that these
additional representations do not lead to any destabilization of
the gauge hierarchy. 

An alternative possibility is the operator 
${\bf 16}_i {\bf 16}_j T_1 P^{\dag2}$,
which can be obtained by introducing the fields ${\bf 16}'(- \frac{1}{2}
^{+-})$ and ${\bf \overline{16}'}(\frac{1}{2}^{++})$.  Again, the
addition of these fermions does not destabilize the gauge hierarchy.
The subject of suitable higher-order diagrams for the vanishing first and
second generation elements of the mass matrices in (8) - (10) and (12)
is under investigation, and the results will be reported elsewhere.

We have calculated the effect of the superheavy quarks and leptons
on the running of the gauge couplings. Defining $\epsilon_3 
\equiv [\alpha_3 (M_G) - \tilde{\alpha}_G]/\tilde{\alpha}_G$, as in
\cite{l-r}, we find that the quarks and leptons contribute $-0.004$.
Though this is in the right direction to improve the fit to the data,
it is too small to be significant as the discrepancy is on the 
order of 2 or 3 $\%$ in SUSY GUTs \cite{l-r}.

\section{Summary}

\qquad We have thus been able to show that it is possible to construct
a realistic set of mass matrices for the quarks and leptons which makes
use of precisely the Higgs fields necessary to solve the doublet-triplet
splitting problem in the $SO(10)$ framework: one {\bf 45} adjoint Higgs 
with 
its VEV pointing in the $B - L$ direction; two pairs of ${\bf 16 + 
\overline{16}}$ spinor Higgs, one of which gets VEV's at the GUT scale 
while 
the {\bf 16} of the other develops an electroweak-breaking VEV in the 
$SU(5)$ 
${\bf \overline{5}}$ direction; and a pair of {\bf 10} vector Higgs, one 
of 
which
develops a pair of electroweak-breaking doublets.  The ${\bf \overline{5}}$
({\bf 16}) and ${\bf \overline{5}}$({\bf 10}) mix with the mixing angle 
possibly 
serving to achieve a small $m^0_b/m^0_t$ ratio without necessitating a large
$\tan \beta$.  Just one pair of vectorlike superheavy fermions in the 
${\bf 16 + \overline{16}}$ spinor and {\bf 10 + 10'} vector representations
are required to generate masses for the second and third generations of 
quarks and leptons.  Higher-order radiative corrections will give masses
to the first generation fermions and are under study. 

An interesting consequence of the incorporation of the Georgi-Jarlskog
factor of three in the quark and charged lepton mass matrices is the 
prediction of sizable $\nu_{\mu} - \nu_{\tau}$ mixing in the neutrino
sector without the imposition of a special texture for the right-handed
Majorana matrix.  This has a direct bearing on the large $\mu -
\tau$ neutrino mixing observed with atmospheric neutrinos and in future
long-baseline experiments.\\  

We thank Keith Dienes and Jens Erler for helpful discussions about the
possibilities for multiple adjoint Higgs fields in superstring theory.
The research of SMB was supported in part by Department of Energy Grant 
Number DE FG02 91 ER 40626 A007.  One of us (CHA) thanks the Fermilab 
Theoretical Physics Department for its kind hospitality where much of 
his work was carried out.  Fermilab is operated by Universities Research
Association Inc. under contract with the Department of Energy.

\vspace*{0.25in}

\noindent
{\Large \bf Figure Captions:}

\noindent
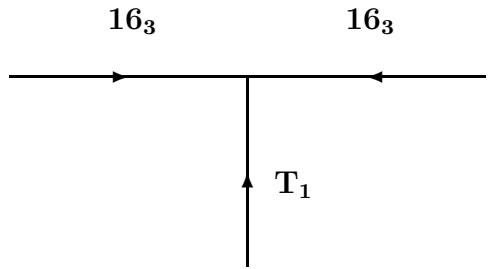
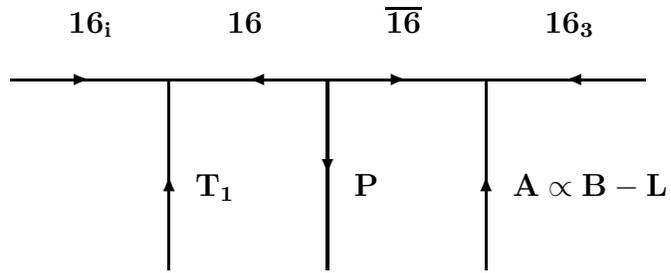
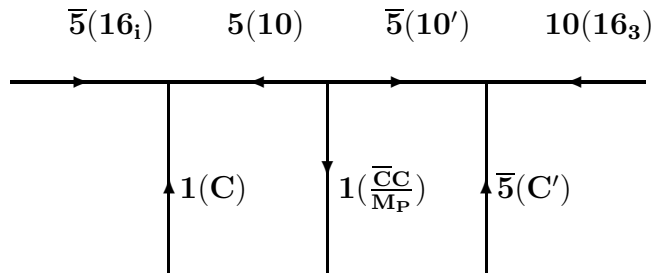
{\bf Fig. 1:} Diagrams that generate the entries in the quark and
lepton mass matrices shown in Eqs. (8) - (10).  (a) The 33 elements denoted
``E". (b) The 23 and 32 elements denoted ``G". Note that because of
the VEV of $A$ they are proportional to the $SO(10)$ generator $B-L$.
(c) The asymmetric entries denoted ``$G$" and ``$G'$" arise from these
diagrams. That they do not contribute to the up quark masses, and
contribute asymmetrically to the down quark and lepton mass matrices,
are consequences of the fact that the $SO(10)$ ${\bf 10}$'s contain
${\bf \overline{5}}$ but not ${\bf 10}$ of $SU(5)$. 

\vspace{0.1cm}

\noindent
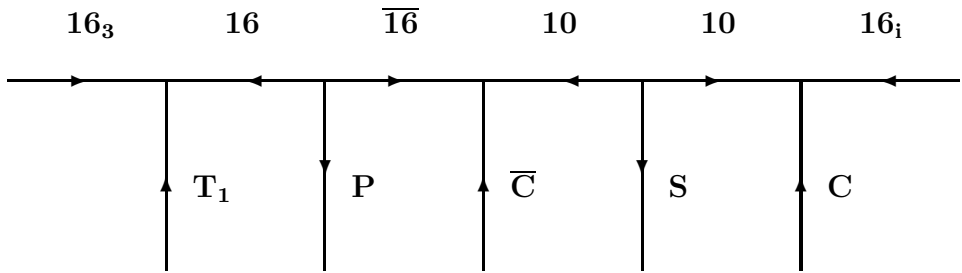
{\bf Fig. 2:} A diagram that could generate the ``$E$" and ``$E'$"
entries of the mass matrices in an alternative version of the model.
However, this version has an unstable gauge hierarchy. Thus the
diagram in Fig. 1(c) is necessary, implying that $C'$ must break
the weak interactions.

\vspace{0.1cm}

\noindent
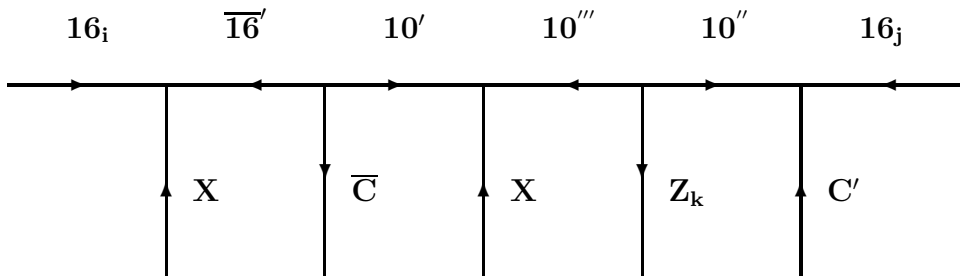
{\bf Fig. 3:} A diagram that can generate small masses for the first
generation quarks and leptons. 
\newpage
\begin{picture}(360,216)
\thicklines
\put(90,144){\vector(1,0){45}}
\put(135,144){\line(1,0){45}}
\put(270,144){\vector(-1,0){45}}
\put(180,144){\line(1,0){45}}
\put(180,72){\vector(0,1){36}}
\put(180,108){\line(0,1){36}}
\put(127,162){${\bf 16_3}$}
\put(217,162){${\bf 16_3}$}
\put(190,100){${\bf T_1}$}
\put(162,36){{\bf Fig. 1(a)}}
\end{picture}

\vspace*{-0.2in}

\begin{picture}(360,216)
\thicklines
\put(60,144){\vector(1,0){30}}
\put(90,144){\line(1,0){30}}
\put(120,144){\line(1,0){30}}
\put(180,144){\vector(-1,0){30}}
\put(180,144){\vector(1,0){30}}
\put(210,144){\line(1,0){30}}
\put(240,144){\line(1,0){30}}
\put(300,144){\vector(-1,0){30}}
\put(120,72){\vector(0,1){36}}
\put(120,108){\line(0,1){36}}
\put(180,72){\line(0,1){36}}
\put(180,144){\vector(0,-1){36}}
\put(240,72){\vector(0,1){36}}
\put(240,108){\line(0,1){36}}
\put(82,162){${\bf 16_i}$}
\put(142,162){${\bf 16}$}
\put(202,162){${\bf \overline{16}}$}
\put(262,162){${\bf 16_3}$}
\put(130,100){${\bf T_1}$}
\put(190,100){${\bf P}$}
\put(250,100){${\bf A \propto B-L}$}
\put(162,36){{\bf Fig. 1(b)}}
\end{picture}

\vspace*{-0.2in}

\begin{picture}(360,216)
\thicklines
\put(60,144){\vector(1,0){30}}
\put(90,144){\line(1,0){30}}
\put(120,144){\line(1,0){30}}
\put(180,144){\vector(-1,0){30}}
\put(180,144){\vector(1,0){30}}
\put(210,144){\line(1,0){30}}
\put(240,144){\line(1,0){30}}
\put(300,144){\vector(-1,0){30}}
\put(120,72){\vector(0,1){36}}
\put(120,108){\line(0,1){36}}
\put(180,72){\line(0,1){36}}
\put(180,144){\vector(0,-1){36}}
\put(240,72){\vector(0,1){36}}
\put(240,108){\line(0,1){36}}
\put(82,162){${\bf \overline{5}(16_i)}$}
\put(142,162){${\bf 5(10)}$}
\put(202,162){${\bf \overline{5}(10')}$}
\put(262,162){${\bf 10(16_3)}$}
\put(124,100){${\bf 1(C)}$}
\put(184,100){${\bf 1(\frac{\overline{C} C}{M_P})}$}
\put(244,100){${\bf \overline{5}(C')}$}
\put(162,36){{\bf Fig. 1(c)}}
\end{picture}

\newpage
\begin{picture}(360,216)
\thicklines
\put(0,144){\vector(1,0){30}}
\put(30,144){\line(1,0){30}}
\put(60,144){\line(1,0){30}}
\put(120,144){\vector(-1,0){30}}
\put(120,144){\vector(1,0){30}}
\put(150,144){\line(1,0){30}}
\put(180,144){\line(1,0){30}}
\put(240,144){\vector(-1,0){30}}
\put(240,144){\vector(1,0){30}}
\put(270,144){\line(1,0){30}}
\put(300,144){\line(1,0){30}}
\put(360,144){\vector(-1,0){30}}
\put(60,72){\vector(0,1){36}}
\put(60,108){\line(0,1){36}}
\put(120,72){\line(0,1){36}}
\put(120,144){\vector(0,-1){36}}
\put(180,72){\vector(0,1){36}}
\put(180,108){\line(0,1){36}}
\put(240,72){\line(0,1){36}}
\put(240,144){\vector(0,-1){36}}
\put(300,72){\vector(0,1){36}}
\put(300,108){\line(0,1){36}}
\put(22,162){${\bf 16_3}$}
\put(82,162){${\bf 16}$}
\put(142,162){${\bf \overline{16}}$}
\put(202,162){${\bf 10}$}
\put(262,162){${\bf 10}$}
\put(322,162){${\bf 16_i}$}
\put(70,100){${\bf T_1}$}
\put(130,100){${\bf P}$}
\put(190,100){${\bf \overline{C}}$}
\put(250,100){${\bf S}$}
\put(310,100){${\bf C}$}
\put(162,36){{\bf Fig. 2}}
\end{picture}

\begin{picture}(360,216)
\thicklines
\put(0,144){\vector(1,0){30}}
\put(30,144){\line(1,0){30}}
\put(60,144){\line(1,0){30}}
\put(120,144){\vector(-1,0){30}}
\put(120,144){\vector(1,0){30}}
\put(150,144){\line(1,0){30}}
\put(180,144){\line(1,0){30}}
\put(240,144){\vector(-1,0){30}}
\put(240,144){\vector(1,0){30}}
\put(270,144){\line(1,0){30}}
\put(300,144){\line(1,0){30}}
\put(360,144){\vector(-1,0){30}}
\put(60,72){\vector(0,1){36}}
\put(60,108){\line(0,1){36}}
\put(120,72){\line(0,1){36}}
\put(120,144){\vector(0,-1){36}}
\put(180,72){\vector(0,1){36}}
\put(180,108){\line(0,1){36}}
\put(240,72){\line(0,1){36}}
\put(240,144){\vector(0,-1){36}}
\put(300,72){\vector(0,1){36}}
\put(300,108){\line(0,1){36}}
\put(22,162){${\bf 16_i}$}
\put(82,162){${\bf \overline{16}'}$}
\put(142,162){${\bf 10'}$}
\put(202,162){${\bf 10^{'''}}$}
\put(262,162){${\bf 10^{''}}$}
\put(322,162){${\bf 16_j}$}
\put(70,100){${\bf X}$}
\put(130,100){${\bf \overline{C}}$}
\put(190,100){${\bf X}$}
\put(250,100){${\bf Z_k}$}
\put(310,100){${\bf C'}$}
\put(162,36){{\bf Fig. 3}}
\end{picture}

\end{document}